\documentclass[shortnote,twocolumn]{jpsj3}

\usepackage{graphicx}
\usepackage{dcolumn}
\usepackage{bm}
\usepackage{color}
\usepackage{ulem}
\usepackage{txfonts}

\def\kB{{k_{\rm B}}}

\def\eq.#1{Eq.~(\ref{#1})}
\def\eqs.#1{Eqs.~(\ref{#1})}

\def\refeq#1{(\ref{#1})}

\def\Hc2{{H_{\rm c2}}}
\def\difHc2{{H'_{\rm c2}}}

\def\*red*#1{{\color{red}{#1}}}

\newcommand\Equation[2]{
\begin{equation}\label{#1} 
#2
\end{equation}
}

\title{
$d'Q = TdS$ for 
Infinitesimal Irreversible Processes: 
On the Differential Form \\ 
of the Clausius Inequality 
}

\author{Hiroshi Shimahara}

\inst{
Graduate School of Advanced Science and Engineering, Hiroshima University, 
Higashi-Hiroshima 739-8530, Japan 
}

\recdate{\today}

\abst{
Conventionally, the differential form of the Clausius inequality 
is adopted for infinitesimal irreversible processes. 
We show 
that the correct first-order form 
for infinitesimal quasi-static irreversible processes 
is the equality $d'Q=TdS$ rather than $d'Q < TdS$. 
This provides a purely thermodynamic proof 
that internal entropy production is second-order or higher 
and confirms 
that entropy remains a well-defined state quantity 
even when quasi-static irreversible processes are included, 
without resorting to the Boltzmann relation. 
}

\begin{document}
\sloppy
\maketitle

In phenomenological thermodynamics, 
apart from the statistical interpretation 
via the Boltzmann relation $S = \kB \ln W$, 
entropy $S$ is usually defined through the relation 
\Equation{eq:DQ=TDS} 
{
     \Delta Q = T \Delta S 
     }
for a sufficiently small reversible process, 
where $\Delta Q$ is the heat exchanged, 
$\Delta S$ is the change in entropy, 
and $T$ is the temperature.\cite{Cla1865,Car09} 
It is generally understood that the process must be reversible, 
since the proof that entropy is a state function 
typically relies on the analysis of reversible cycles. 
In this relation, $T$ on the right-hand side represents 
the temperature of the system during the course of the small process, 
with variations in $T$ contributing only to higher-order terms. 
Hence, 
the relation \eq.{eq:DQ=TDS} is originally understood as 
valid only to first order in the small variations; 
as a mathematical idealization, 
it is more precisely expressed in the infinitesimal form 
\Equation{eq:dQ=dDS_infinitesimal} 
{
     d'Q = TdS . 
     }
Here, 
$d'Q$ is the inexact differential associated with heat exchange 
and $dS$ is the exact differential of entropy. 
Integrating \eq.{eq:dQ=dDS_infinitesimal} 
along an arbitrary reversible path $C_{\rm r}$ 
connecting the initial state A to the final state B 
leads to 
\Equation{eq:Sdef_in_integral}
{
     S({\rm B}) - S({\rm A}) 
     = \int_{C_{\rm r}} \frac{d'Q_{\rm r}}{T} . 
     }
As noted above, 
the second law of thermodynamics, 
applied to reversible cycles, 
ensures the path independence of this integral, 
which implies that entropy is a state quantity.

In this Short Note, 
we consider irreversible processes as well as reversible processes, 
but restrict ourselves to quasi-static processes,\cite{Note1} 
so that bulk thermodynamic quantities, 
such as the temperature $T$, remain well defined. 
The second law of thermodynamics also leads to 
the Clausius inequality\cite{Cla1854,Cla1865} 
\Equation{eq:int_r_int_ir} 
{
     \int_{\rm A}^{\rm B} \frac{d'Q_{\rm r}}{T} 
     > 
     \int_{\rm A}^{\rm B} \frac{d'Q_{\rm ir}}{T_{\rm bath}} , 
     }
where 
$d'Q_{\rm r}$ and $d'Q_{\rm ir}$ denote 
the infinitesimal heat exchanges along, respectively, 
arbitrary reversible and irreversible processes 
connecting states A and B. 
The left-hand side of \eq.{eq:int_r_int_ir} 
is equal to $S({\rm B}) - S({\rm A})$ 
as defined in \eq.{eq:Sdef_in_integral}. 
In what follows, 
both A and B are assumed to be quasi-equilibrium states 
in the sense that bulk thermodynamic variables are well defined. 
The temperature $T_{\rm bath}$ appearing in the integral 
for the irreversible process denotes the temperature of the heat source 
with which the system is in contact at each stage of the process 
between states A and B. 
When the processes are sufficiently small, 
\eq.{eq:int_r_int_ir} may appear, 
under the implicit assumption that $T_{\rm bath} \approx T$, 
to reduce to 
the conventional interpretation of the Clausius inequality, 
\Equation{eq:conv_Clausius}
{
     \Delta S 
     = \frac{\Delta Q_{\rm r}}{T} 
     > \frac{\Delta Q_{\rm ir}}{T} , 
     } 
or, equivalently, 
\Equation{eq:2ndLaw_small} 
{
     \Delta Q_{\rm ir} < T \Delta S , 
     }
which appears to contrast with \eq.{eq:DQ=TDS} for reversible processes. 
In fact, standard textbooks 
[e.g., Refs.~\citen{Kub68, Zem97, Kon14}] 
conventionally 
extend the Clausius inequality for finite processes 
to infinitesimal form, such as $d'Q \le TdS$, 
in which the strict inequality $<$ is understood to hold for 
irreversible processes. 
However, 
when the processes are so small that 
the definition \eq.{eq:DQ=TDS} applies, 
or when they are infinitesimal, 
these extended inequalities are 
not consistent with the first-order (infinitesimal) formulation. 
As we show below, the correct first-order relations are 
\Equation{eq:corr_Clausius_dif}
{
        \Delta Q_{\rm ir} = T \Delta S 
        \qquad \mbox{and} \qquad
        d'Q_{\rm ir} = T dS . 
     }

Applying the first law of thermodynamics, 
$\Delta E = W + Q$, 
to the sufficiently small reversible and irreversible processes 
considered here, 
we obtain, to first order in the small variations, 
\Equation{eq:DE=DW+DQ} 
{
     \begin{split} 
     \Delta E & = \Delta W_{\rm r} + \Delta Q_{\rm r} \\ 
     \Delta E & = \Delta W_{\rm ir} + \Delta Q_{\rm ir} , 
     \end{split} 
     }
where 
$\Delta W_{\rm r}$ and 
$\Delta W_{\rm ir}$ 
denote the work done on the system 
in the reversible and irreversible processes, respectively. 
Since the two processes connect 
the same initial and final quasi-equilibrium states, 
the same $\Delta E$ appears in both equations. 
The work in each process is a purely mechanical quantity. 
As with $\Delta E$, 
the volume change $\Delta V$ is identical for both processes. 
Under the present quasi-static assumption, 
the pressure can be treated as a well-defined bulk variable. 
The pressures $p_{\rm r}$ and $p_{\rm ir}$ 
along the reversible and irreversible processes 
may vary and differ. 
However, 
for the two processes connecting the same initial and final states, 
such variations and differences are small quantities. 
Therefore, 
since the pressures contribute to the work only through 
a product with the small volume change $\Delta V$, 
the variations and differences 
do not contribute at first order, but only at higher order. 
Hence, we can define a common representative value $p$ 
of the pressures, 
which leads to the first-order relations 
$\Delta W_{\rm r} = - p \Delta V$ 
and $\Delta W_{\rm ir} = - p \Delta V$. 
Consequently, we obtain, to first order, 
\Equation{eq:DQr=DQir}
{ 
      \Delta Q_{\rm r} = \Delta E + p \Delta V = \Delta Q_{\rm ir} , 
      } 
and hence, in the infinitesimal limit, 
the corresponding 1-form relation 
\Equation{eq:dQr=dQir_infini}
{ 
     d'Q_{\rm r} = d'Q_{\rm ir} . 
     } 
It should be emphasized that 
these equations are obtained 
without assuming $T_{\rm bath} \approx T$; 
rather, 
they follow directly from the first law 
and the first-order expression for the mechanical work.

Hence, combining these results with 
the definition of entropy through 
$\Delta Q_{\rm r} = T \Delta S$ and 
$d'Q_{\rm r} = T dS$ for reversible processes, 
the first-order relations in \eq.{eq:corr_Clausius_dif} follow. 
Thus, although the relations 
Eqs.~\refeq{eq:DQ=TDS} and \refeq{eq:dQ=dDS_infinitesimal} 
are usually considered only for reversible processes, 
they also hold for the irreversible processes considered here, 
once one recognizes 
that \eq.{eq:DQ=TDS} is intrinsically a first-order relation 
and 
that \eq.{eq:dQ=dDS_infinitesimal} is the corresponding 1-form relation. 
In particular, the first law of thermodynamics 
in the form 
\Equation{eq:firstlaw}
{
     \Delta E = - p \Delta V + T \Delta S 
     } 
can be used not only for reversible processes 
but also for irreversible processes, 
provided that the processes are quasi-static, 
because \eq.{eq:firstlaw} is itself a first-order relation.

The present result in \eq.{eq:corr_Clausius_dif} 
provides a simple proof 
that the internal entropy production $S_{\rm prod}$ 
for small quasi-static irreversible processes 
is of second order or higher. 
For a finite irreversible process under consideration, 
$S_{\rm prod}$ is commonly defined by 
\Equation{eq:entropy_prod_def}
{
     S_{\rm prod} = S({\rm B}) - S({\rm A}) 
        - \int_{\rm A}^{\rm B} \frac{d'Q_{\rm ir}}{T} , 
     }
where $T$ denotes the temperature of the system. 
The vanishing of the difference $\Delta S - \Delta Q_{\rm ir}/T$ 
to first order 
immediately implies that, 
for a sufficiently small process, 
the internal entropy production 
\Equation{eq:difference}
{
     S_{\rm prod}
     = 
     \Delta S - \frac{\Delta Q_{\rm ir}}{T} , 
     } 
is of second order or higher. 
Although this fact is well known,\cite{Ons31a,Ons31b,Pri78,Vel11,Lan21} 
we have derived it directly from basic thermodynamic principles, 
without involving microscopic assumptions 
or detailed statistical-mechanical calculations 
based on the Boltzmann relation.

As a direct consequence of the infinitesimal relation 
$d'Q_{\rm r} = d'Q_{\rm ir}$, 
one obtains, by integration along the corresponding finite paths 
$C_{\rm r}$ and $C_{\rm ir}$, 
\Equation{eq:integral_dQ}
{
     Q(C_{\rm r}) 
       \equiv \int_{C_{\rm r}} d'Q_{\rm r} 
            = \int_{C_{\rm ir}} d'Q_{\rm ir} 
       \equiv 
     Q(C_{\rm ir}) . 
     }
By construction, 
the paths $C_{\rm r}$ and $C_{\rm ir}$ are 
geometrically identical in the thermodynamic state space, 
while the processes along the paths 
are reversible and irreversible, respectively. 
This result may appear to follow directly from 
the first law of thermodynamics, 
$\Delta E = W + Q$, 
without the intermediate steps used here, 
i.e., first passing to the differential form 
and then integrating it; 
however, 
it should be noted that, for finite paths, 
$W_{\rm r} = W_{\rm ir}$ is not generally guaranteed, 
but requires that 
the states be quasi-equilibrium states 
and the processes be quasi-static.

Another direct consequence of $d'Q_{\rm r} = d'Q_{\rm ir}$ is 
\Equation{eq:integral_dQ_T}
{
     S({\rm B}) - S({\rm A}) 
       = \int_{C_{\rm r}}
         \frac{d'Q_{\rm r} }{T}
       = \int_{C_{\rm ir}}
         \frac{d'Q_{\rm ir} }{T} . 
     }
This relation implies that the {\it internal} entropy production defined 
by \eq.{eq:entropy_prod_def} vanishes 
provided that the process is quasi-static.\cite{Note2}
This may be one source of the common misconception 
that a quasi-static process is necessarily reversible. 
In fact, 
the thermal irreversibility in such processes 
originates from the heat transfer across the boundary 
between the system and the heat bath (heat source), 
driven by the temperature difference $T \ne T_{\rm bath}$. 
This yields the {\it total} entropy production 
$S_{\rm prod}^{(\rm tot)}$, 
which is readily verified from the second law of thermodynamics 
to be positive: 
\Equation{eq:entropy_prod_tot}
{
     S_{\rm prod}^{(\rm tot)} 
     =   \int_{C_{\rm ir}} \frac{d'Q_{\rm ir}}{T}
       - \int_{C_{\rm ir}} \frac{d'Q_{\rm ir}}{T_{\rm bath}}
     > 0 . 
     }

The relation \eq.{eq:integral_dQ_T}
also means that 
when the states A and B are connected by 
an arbitrary irreversible path $C_{\rm ir}$ 
within the present assumptions, 
the value of the integral along $C_{\rm ir}$ 
is the same as 
that along a corresponding reversible path $C_{\rm r}$.\cite{Note3}
As discussed after \eq.{eq:Sdef_in_integral}, 
the standard thermodynamic construction defines entropy 
as a state quantity 
by proving the path independence of \eq.{eq:Sdef_in_integral} 
within the class of reversible paths, 
which does not by itself imply path independence 
for irreversible paths. 
The relation \eq.{eq:integral_dQ_T} 
extends this path independence to the present class of 
quasi-static irreversible paths, 
on the basis of purely thermodynamic considerations, 
without resorting to the Boltzmann relation.

In summary, 
through simple arguments directly based on thermodynamic principles, 
we have shown that 
the correct first-order relations for quasi-static 
irreversible processes are 
$\Delta Q_{\rm ir} = T \Delta S$ 
and 
$d'Q_{\rm ir} = TdS$, 
from which it follows that 
the internal entropy production is of second order or higher. 
We have also clarified that 
the form of the first law of thermodynamics in \eq.{eq:firstlaw} 
remains valid for quasi-static irreversible processes, 
and have provided a purely thermodynamic proof, 
independent of the Boltzmann relation, 
that the path independence of the entropy integral extends 
to quasi-static irreversible paths.



\end{document}